\documentclass[amsfonts,amsmath]{PoS}
\usepackage{amsfonts,amsmath}

\graphicspath{{./Figures/}}

\newcommand{\noi}{\noindent}
\newcommand{\beq}{\begin{equation}}
\newcommand{\eeq}{\end{equation}}
\newcommand{\bea}{\begin{eqnarray}}
\newcommand{\eea}{\end{eqnarray}}

\newcommand{\Fig}[1]{Fig.~\ref{#1}}
\newcommand{\Tab}[1]{Table~\ref{#1}}
\newcommand{\Sec}[1]{Section~\ref{#1}}

\newcommand{\aleq}{\mbox{}_{\textstyle \sim}^{\textstyle < }}
\newcommand{\ageq}{\mbox{}_{\textstyle \sim}^{\textstyle > }}

\newcommand{\Nt}{N_{\tau}}
\newcommand{\Ns}{N_{\sigma}}
\newcommand{\cV}{\mathcal{V}}
\newcommand{\cB}{\mathcal{B}}
\newcommand{\cE}{\mathcal{E}}

\title{Two-colour QCD at non-zero temperature \\ 
       in the presence of a strong magnetic field}

\ShortTitle{Two-colour QCD with magnetic field}

\author{\speaker{M. M\"uller-Preussker}\,%
\thanks{MMP expresses his gratitude to the organizers for having 
        been invited to this interesting workshop and 
        $~~~~~~~~~$ to the ECT* staff for kind hospitality.}~~, 
        B. Petersson, A. Schreiber \\
        Humboldt-Universit\"at zu Berlin, Institut f\"ur Physik, 
        12489 Berlin, Germany \\
        E-mail: \email{ mmp@physik.hu-berlin.de, 
                       bengt.petersson@physik.hu-berlin.de,
                       alexander.schreiber@physik.hu-berlin.de}}
\author{E.-M. Ilgenfritz\\
       Joint Institute for Nuclear Research, VBLHEP and BLTP, 
       141980 Dubna, Russia \\
        E-mail: \email{Michael.Ilgenfritz@sunse.jinr.ru}}
\author{M. Kalinowski \\ 
        Goethe-Universit\"at Frankfurt am Main, Institut f\"ur Theoretische Physik,
        60438 Frankfurt am Main, Germany \\
        E-mail: \email{kalinowm@th.physik.uni-frankfurt.de}}

\leftline{\parbox{11cm}{\large\rm HU-EP-14/03}}
\vspace*{-0.5cm}

\abstract{
In this talk we report on our study of two-colour lattice QCD 
with $N_f=4$ staggered fermion degrees of freedom with equal electric 
charge $q$ in a homogeneous magnetic field $B$ at non-zero temperature $T$. 
We find indications for a non-monotonic behaviour of the critical temperature 
as a function of the magnetic field strength and, as a consequence, 
for the occurence of {\it inverse magnetic catalysis} within the transition
region for magnetic fields in the range $0 \le qB \aleq 0.7~\mathrm{GeV}^2$  .        
}

\FullConference{
QCD-TNT-III-From quarks and gluons to hadronic matter: 
A bridge too far?,\\ 2-6 September, 2013\\
European Centre for Theoretical Studies in Nuclear Physics and 
Related Areas (ECT*), Villazzano, Trento (Italy)}

\begin{document}

\section{Introduction}
The behaviour of hadronic matter under the influence of a strong
magnetic field $B$ has recently been widely discussed because of its 
relevance for non-central heavy ion collisions. In such collisions 
there will be two lumps of spectators moving in opposite directions. 
They give rise to a magnetic field perpendicular to the reaction plane.
It can be shown that the magnetic field is so strong that its 
consequences cannot be studied perturbatively \cite{Kharzeev:2007jp,
Skokov:2009qp,McLerran:2013hla}. 

The influence of an external magnetic field on hadronic matter at zero temperature
has been studied by various authors, e.g. within the Nambu-Jona-Lasinio 
\cite{Klevansky:1989vi} or in the chiral model \cite{Shushpanov:1997sf}. 
The general result is that the magnetic field induces an increase of 
the chiral condensate. This was called {\it magnetic catalysis} in 
\cite{Gusynin:1994re} and claimed to be essentially model independent.
For a recent review see also \cite{Shovkovy:2012zn}. The model calculations have 
been extended also to finite temperature $T$, in order to study the phase diagram of 
strongly interacting matter in a constant magnetic field. The critical 
temperature of the chiral phase transition rises in most of the calculations 
\cite{Klimenko:1992ch,Agasian:2001hv}. 
But there are also claims that the chiral and the deconfinement phase transitions 
may split, and the critical temperature of the latter decreases with the magnetic 
field strength \cite{Agasian:2008tb,Mizher:2010zb,Andersen:2012jf,Orlovsky:2013xxa}.

A couple of years ago several groups have started to investigate the problem 
through {\it ab initio} lattice simulations of QCD and QCD-like theories with
a homogenous magnetic background field. 
The pioneering work - employing quenched $SU(2)$ - was performed by 
M. Polikarpov $\dagger$ and his collaborators \cite{Buividovich:2008wf,
Buividovich:2009ih}. Later on, a few groups have performed investigations in full 
lattice QCD (in Pisa \cite{D'Elia:2010nq,D'Elia:2011zu},
in Regensburg \cite{Bali:2011qj,Bali:2012zg,Bali:2013esa,Bruckmann:2013oba},
see also \cite{Levkova:2013qda} and very recently \cite{Bornyakov:2013eya}). 
All groups observe magnetic catalysis for temperatures in the confined 
phase. In the transition or (better) crossover region the Regensburg group 
reported what they call {\it inverse magnetic catalysis},  
i.e. the chiral condensate and thus the (pseudo-)critical temperature 
decrease with increasing magnetic field strength. 
A nice recent review of the lattice results for QCD and QCD-like theories 
in external fields can be found in Ref. \cite{D'Elia:2012tr}.

In this talk we report on our two-colour QCD investigations 
\cite{Ilgenfritz:2012fw,Ilgenfritz:2013ara} with $N_f=4$ 
flavour fermion degrees of freedom with equal electric charges 
(avoiding any ``rooting'' of the fermionic determinant). 
In this case a first order finite temperature transition 
\cite{Pisarski:1983ms} can be expected in contrast to the observed smooth 
crossover for $N_f=2$ or $2+1$ at small but non-vanishing $u$-, $d$-quark 
masses. Although our model is not QCD, the chiral properties are quite similar. 
Furthermore, investigations of the dynamical $SU(2)$ theory are of considerable 
interest, because they can be extended to finite chemical potential 
without a sign problem. 

In \Sec{sec:setup} we specify the action and our observables as well
as the setup for our simulations. 
In \Sec{sec:results} we discuss the temperature and magnetic field dependence
of the Polyakov loop and chiral condensate as well as of their respective
susceptibilities. Moreover, we provide results of a recent fixed-scale study
at smaller quark mass. In \Sec{sec:conclusions} we provide a conjecture 
about the $B - T$ phase diagram and on the occurence of (inverse) magnetic 
catalysis.     

\section{Setup of the lattice investigation}
\label{sec:setup}

We introduce a lattice of four-dimensional size $\cV\equiv \Nt\times \Ns^3$
with a spacing unit $a$. The physical volume and temperature are 
$V = (a\Ns)^3$ and $T= 1/ (a\Nt)$, respectively.
On the links $n \to n+\hat{\mu}$ the group elements 
$U_{\mu}(n)\in SU(2)$, $\mu = 1,2,3,4$ are defined. Periodic boundary 
conditions are assumed. We employ the standard Wilson plaquette action  
\beq
S_G= \beta\cV \sum_{\mu<\nu}P_{\mu\nu}, \qquad
P_{\mu\nu} = \frac{1}{\cV}\sum_n(\frac{1}{2}Tr\left({\bf 1}-U_{\mu\nu}(n)\right))
\eeq
\noi
with $U_{\mu\nu}(n)$ denoting the $\mu\nu$-plaquette matrix at site $n$.
For the fermion part of the action, we use staggered Grassmann variables 
$\bar{\chi}_n$ and $\chi_n$ transforming with the fundamental 
representation of the gauge group $SU(2)$.
For simplicity the four flavour degrees of freedom are assumed to carry equal 
electric charges $q$ allowing to interact with an external magnetic field $B$. 
The boundary conditions of the fermionic fields are (anti-) periodic
in the space (time) directions. In the absence of a magnetic field the 
fermionic part of the action reads
\beq
S_F  =  a^3\sum_{n,n^{\prime}} \bar{\chi}_n [D_{n,n^{\prime}} 
     + ma\delta_{n,n^{\prime}}]\chi_{n^{\prime}}, \quad
D_{n,n^{\prime}} 
 =  \frac{1}{2}\sum_{\mu}\eta_{\mu}(n)[U_{\mu}(n)\delta_{n+\mu,n^{\prime}}-  
    U_{\mu}^{\dagger}(n-\mu)\delta_{n-\mu,n^{\prime}}], \label{stag}
\eeq
where $m$ is the bare quark mass.
\noi
The $\eta_{\mu}(n)$ are the standard staggered sign factors,
\beq
\eta_1(n) = 1\,, \hspace{1cm}
\eta_{\mu}(n) = (-1)^{\sum_{\nu=1}^{\mu-1} n_{\nu}}\,, \hspace{0.5cm} 
\mu=2,3,4\,.
\eeq
We introduce electromagnetic background potentials into the 
fermion action by new, commuting group elements on the
links, namely $V_{\mu}(n)= e^{i\theta_{\mu}(n)}\in U(1)$.
A constant magnetic background field in the $z \equiv 3$-direction 
penetrating through all the $(x,y) \equiv (1,2)$ -planes of finite size 
$\Ns \times \Ns$ with a constant magnetic flux $\phi =a^2qB$ 
through each plaquette can be realized as follows:
\bea
&~& V_1(n)  =  e^{-i\phi n_2/2} \hspace{0.4cm} (n_1 =1,2,\ldots , \Ns-1)\,, 
\qquad V_2(n)  =  e^{i\phi n_1/2} \hspace{0.5cm} (n_2=1,2,\ldots ,\Ns-1)\,, 
\nonumber \\
&~& V_1(\Ns,n_2,n_3,n_4)  =  e^{-i\phi(\Ns+1) n_2/2}\,, 
\hspace{1.9cm} V_2(n_1,\Ns,n_3,n_4)  =  e^{i\phi(\Ns+1) n_1/2}\,,  \nonumber  \\
&~& V_3(n)=V_4(n)=1\,.
\eea
With periodic boundary conditions the magnetic flux becomes quantized as
$\phi=a^2qB= 2\pi N_b / \Ns^2, ~N_b \in Z.$ 
Because the angle $\phi$ is periodic, the flux is bounded from above 
$\phi < \pi$. One obtains the condition $N_b < \Ns^2/2$. Physically 
reasonable strong fields should then be restricted at least to 
half of this bound, i.e. $N_b \le \Ns^2/4$.

\noi
We introduce the fields $V_{\mu}(\theta)$ into the fermionic action
$S_F(\theta)$ by substituting in Eq. (\ref{stag})
\beq
U_{\mu}(n)  \rightarrow  V_{\mu}(n)U_{\mu}(n)\,, \qquad
U_{\mu}^{\dagger}(n)  \rightarrow  V_{\mu}^{\ast}(n)U^{\dagger}_{\mu}(n)\,. 
\eeq
The partition function in the background field $\theta$ is then given by
\beq
Z(\theta)=\int \prod(d\bar{\chi}(n)d\chi(n) dU_{\mu}(n))e^{-S_G - S_F(\theta)}.
\eeq
The simulation algorithm employed is the usual Hybrid Monte Carlo method,
updated in various respects in order to increase efficiency
(even-odd and mass preconditioning, multiple time scales, Omelyan integrator 
and written in CUDA Fortran for the use on GPU's).  

We have computed the average Polyakov loop $<L>$, which is 
the order parameter for confinement in the limit of infinite quark mass 
\beq
  <L> =  \frac{1}{\Ns^3}\sum_{n_1,n_2,n_3} \frac{1}{2} 
         <\mathrm{Tr} \left( \prod_{n_4=1}^{\Nt} U_4(n_1,n_2,n_3,n_4)\right)>
\eeq
and its susceptibility $\chi_L = \Ns^3(<L^2> - <L>^2)$.
The chiral condensate, which is an exact order parameter in the 
limit of vanishing quark mass, is given by
\beq
a^3<\bar{\chi}\chi>  =  - \frac{1}{\cV}~\frac{1}{4}
~\frac{\partial}{\partial(ma)} \log (Z) 
 =  \frac{1}{\cV}~\frac{1}{4}~<\mathrm{Tr}(D+ma)^{-1}>.
\eeq
In order to locate the phase transition we have used also the disconnected part of 
the susceptibility (called ``chiral susceptibility'' for simplicity),
\bea
 & \chi & = \frac{1}{\Nt\Ns^3}~\frac{1}{4}~\frac{\partial^2}{(\partial(ma))^2} 
\log (Z)
 = \chi_{\mathrm{conn}}+\chi_{\mathrm{disc}}\,, \\
 & \chi_{\mathrm{disc}} & =  \frac{1}{\Nt\Ns^3}~\frac{1}{16}  
                  ( <(\mathrm{Tr}(D+ma)^{-1})^2> - <\mathrm{Tr}(D+ma)^{-1}>^2 ).
\eea
It is important to notice that the mean values defined above are bare quantities
which in principle should be renormalized when comparing with continuum 
expectation values. 

To study the influence of an external magnetic field we have also computed 
the anisotropy in the gluonic action by measuring the average value 
$<P_{\mu\nu}>$ of the non-Abelian plaquette energies for the different 
$\mu-\nu$ planes as a function of the magnetic field strength and of the 
temperature.


A zero-temperature simulation without magnetic field was performed for 
$\beta=1.80$ and for the two mass values $ma=0.0025,~0.01$ on a lattice of 
size $32^3\times 48$ in order to estimate the lattice spacing and the
pion mass (for details see \cite{Ilgenfritz:2012fw,Ilgenfritz:2013ara}).  

For the determination of the lattice spacing $a$ we have computed 
the potential between infinitely heavy quarks. From this we obtained
the Sommer parameter, defined in the continuum by the equation 
\beq
\left. r^2\,\frac{dV}{dr}\right|_{r=r_0}=1.65 \label{r0}\,.
\eeq
Assuming $r_0=0.468(4)$ fm  \cite{Bazavov:2011nk} the lattice spacings 
for $ma=0.0025$ and $ma=0.01$ were obtained (cf. \Tab{tab}). 
Inserting the value of $a$ into the result for the effective 
pseudo-scalar meson mass we obtained the pion mass in physical units
for $ma=0.0025$ and about half of that value for $ma=0.01$ 
as expected from the relation $m_\pi^2 \propto m$ (cf. \Tab{tab}).
\begin{table}[h!]
\centering
\mbox{\scriptsize
 \begin{tabular}{|c|c|c|c|c||c|c|c|c|}
\hline
 $\beta$ & $a m$ & $\Ns$ & $\Nt$ & $N_b^m$ & $R_0$ 
         & $a [\mathrm{fm}]$ & $m_\pi[\mathrm{MeV}]$ & $\sqrt{qB}_{m}[\mathrm{GeV}]$ \\
 \hline
  1.8 & .01  & 16 & 32 & 50  & 2.75(8)  & 0.170(5)  & 330(10)  & 1.29(4)  \\ 
  1.8 & .0025& 32 & 48 & 200 & 2.78(6)  & 0.168(4)  & 175(4)   & 1.30(3)  \\ 
\hline
\end{tabular}
}
\caption{Results for the Sommer scale $R_0$ (in lattice units), the 
lattice spacing $a$, the pion mass $m_\pi$, and   
the magnetic field strength $\sqrt{qB}_m$ for $N_b^m$
flux units \cite{Ilgenfritz:2012fw,Ilgenfritz:2013ara}.}
\label{tab}
\end{table}
As a consequence, for $am=0.01$ we reached a ratio $m_{\pi}/T_c(B=0) \approx 1.7$, 
which is similar to the estimate in \cite{D'Elia:2010nq,D'Elia:2011zu}, 
but higher than that in \cite{Bali:2011qj}. However, for $am=0.0025$ 
we gained a value $m_{\pi}/T_c(B=0) \approx 1.0$ already
interesting for a comparison with the real physical situation.

\section{Results}
\label{sec:results}
\begin{figure*}[tb]
\vspace*{-2cm}
\includegraphics[width=\textwidth]{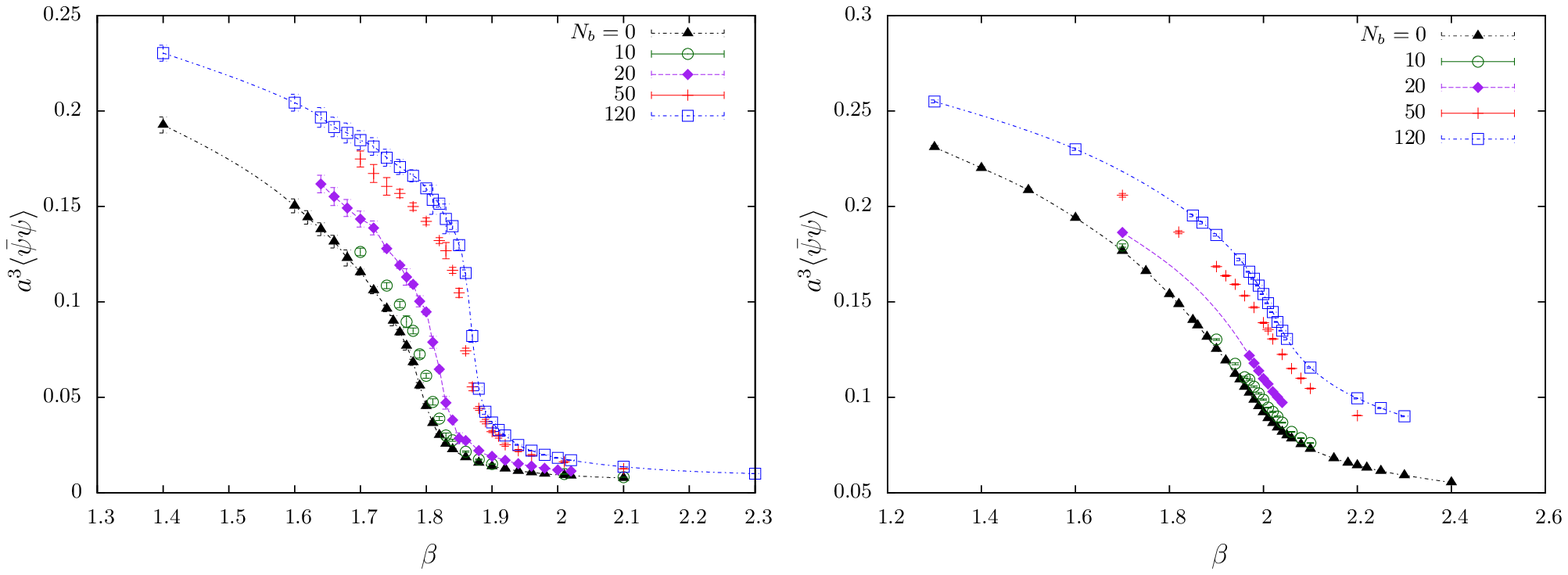}
\vspace*{-15cm}
\caption{Bare chiral condensate $a^3 \langle\bar{\psi}\psi\rangle$ vs. 
$\beta$ for various magnetic fluxes $\phi$ (in flux units), 
$ma=0.01$ (left panel) and $am=0.1$ (right panel), lattice size $16^3 \times 6$. 
Curves are to guide the eye.}
\label{fig:fg1}
\end{figure*}
In \Fig{fig:fg1} we have plotted the bare chiral condensate as a function 
of $\beta$ for a set of numbers of flux quanta and for the bare quark 
mass values $~ma=0.01$ (left panel) and $ma=0.1$ (right panel). 
In both of the cases the chiral condensate increases with rising magnetic 
field for arbitrary fixed $\beta$. For the smaller quark mass we see quite
clearly a transition for all values of the flux quanta $N_b$ under 
consideration. Moreover, the chiral transition seems to move to higher 
temperatures as the magnetic field is increasing. This tendency is in 
agreement with the results in \cite{D'Elia:2010nq,D'Elia:2011zu} but opposite 
to \cite{Bali:2011qj}, where the chiral condensate was seen to decrease 
with the flux $\phi$ in the transition region, leading to a decrease of
the transition temperature. 
\begin{figure*}[tb]
\vspace*{-2cm}
\includegraphics[width=\textwidth]{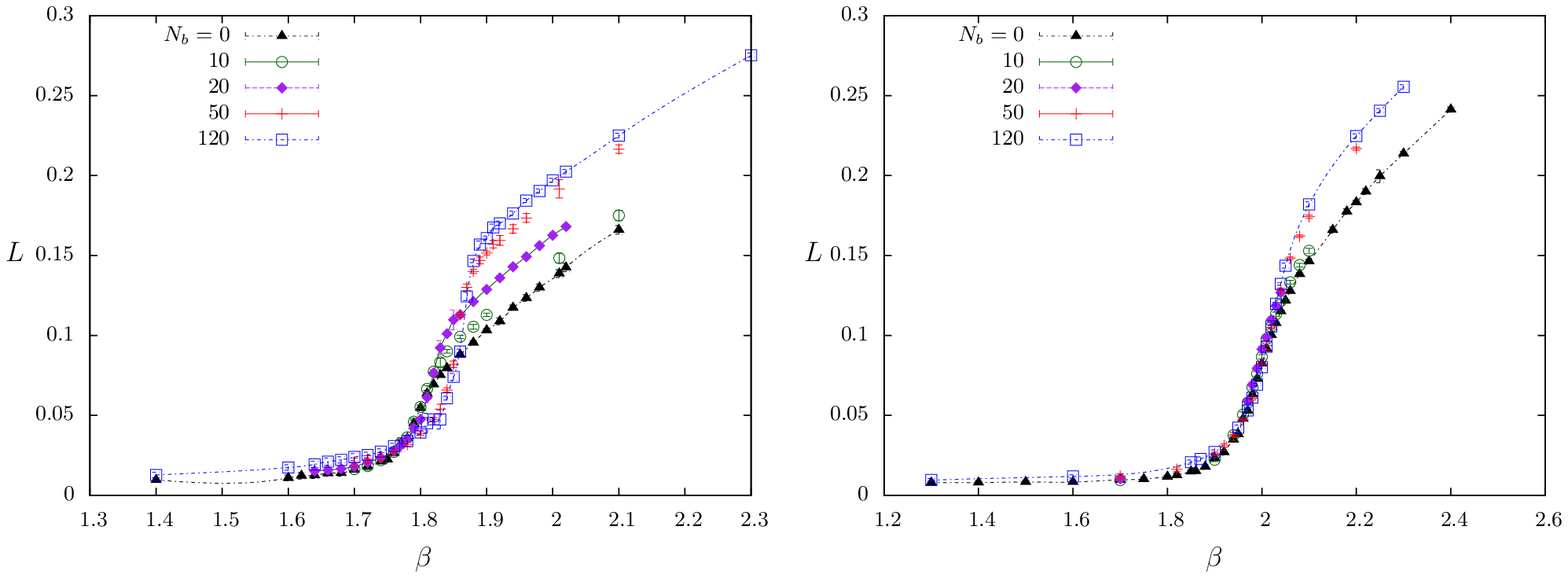}
\vspace*{-15cm}
\caption{Same as for Fig. 1 but for the bare Polyakov loop $<L>$ vs.
$\beta$.  
}
\label{fig:fg2}
\end{figure*}
In \Fig{fig:fg2} the expectation value of the Polyakov loop is shown 
vs. $\beta$ for the same two values of the bare quark mass. 
The transition temperature obviously increases with the quark mass as expected.
At the high quark mass there seems to be only a weak effect of the magnetic
field on the deconfinement temperature. At the smaller mass value we
observe a non-monotonic behaviour with the magnetic field for fixed
$\beta$-values within the transition region.
\begin{figure*}[tb]
\vspace*{-2cm}
\includegraphics[width=\textwidth]{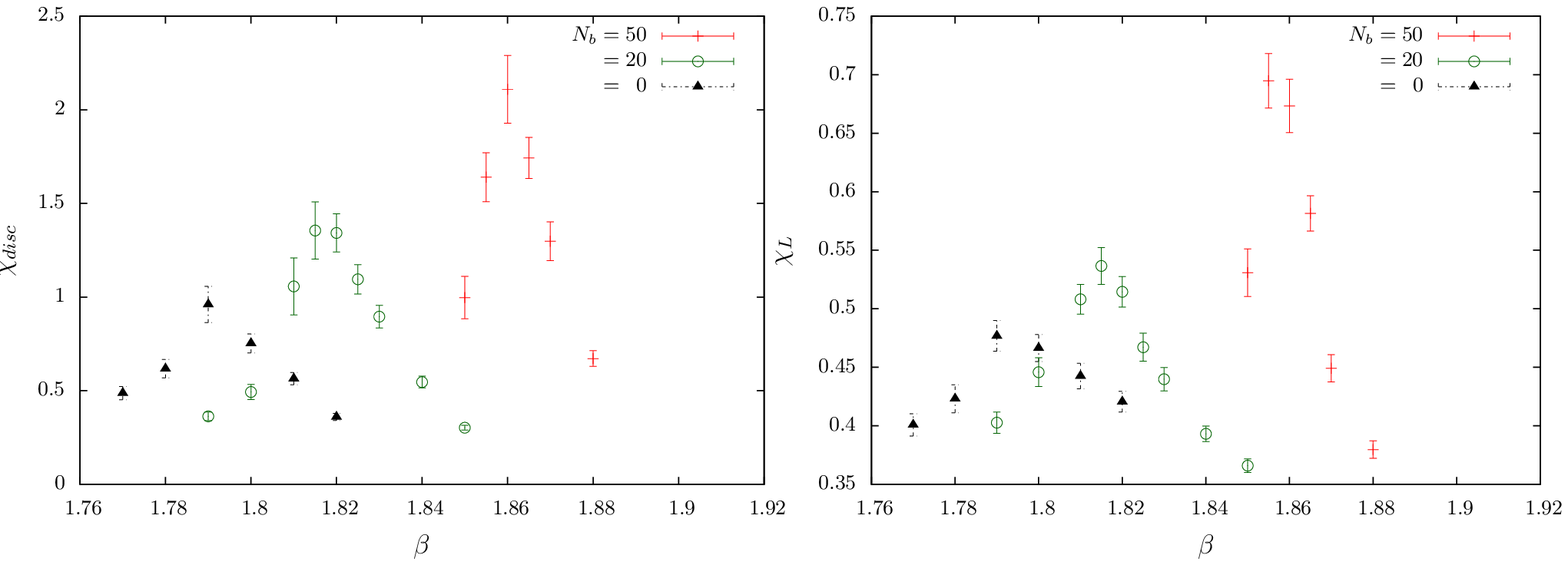}
\vspace*{-15cm}
\caption{The chiral susceptibility $\chi_{disc}$ (left panel) 
and Polyakov loop susceptibility $\chi_{L}$ (right panel)
vs. $\beta$ at $am=0.01$ for magnetic fluxes $N_b$ at lattice 
size $16^3 \times 6$.}
\label{fig:fg3}
\end{figure*}
In \Fig{fig:fg3} we show the chiral susceptibility and the Polyakov loop 
susceptibility for the lower quark mass value $ma=0.01$.
It is clearly seen in the left figure that the chiral transition indeed 
moves to higher temperatures as the magnetic field becomes stronger. 
In the right figure we show the same effect for the Polyakov loop
susceptibility. The maxima of the two susceptibilities turn out to be
at the same value for given magnetic field. 
Thus, there is no sign of a splitting between the chiral and the 
deconfinement transition as it should be expected for a real phase transition. 
However, let us keep in mind that the rise of the temperature $T=1/a(\beta)\Nt$ 
by lowering $a(\beta)$ causes the physical values of the mass $m$ and of the 
magnetic field $qB$ to increase as well, since their values remained fixed 
only in lattice units. At the same time we did not renormalize our observables.
Below we demonstrate how to circumvent these obstacles.    

In order to study the dependence of the chiral condensate on the magnetic 
field strength in the chiral limit, we have looked at 
the behaviour of the chiral condensate as a function of the 
quark mass and of the magnetic field for various $\beta$'s.
Because we now keep $\beta$ fixed we eliminate lattice effects 
coming from the variation of $a$.
We have considered $\beta = 1.70$ (confined phase),
$\beta=1.90$ (transition region), and $\beta=2.10$ (deconfined phase). 
\begin{figure*}[tb]
\vspace*{-2cm}
\includegraphics[width=\textwidth]{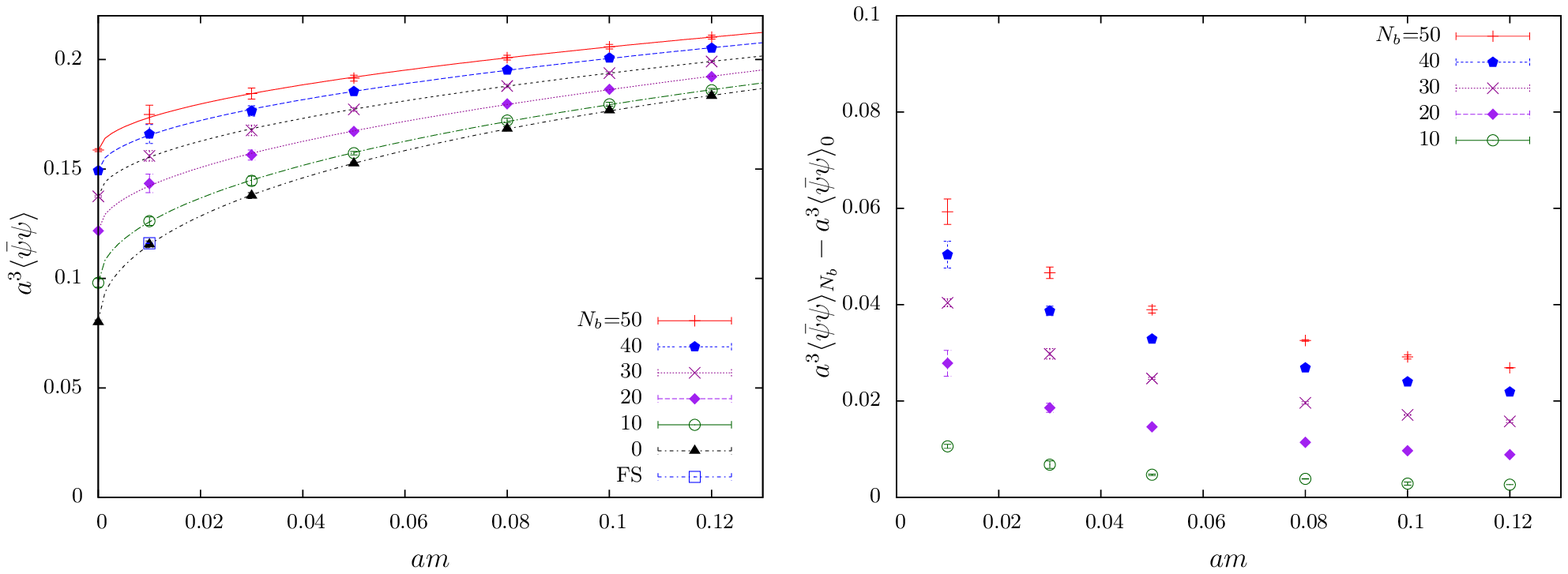}
\vspace*{-15cm}
\caption{Mass dependence of the bare chiral condensate (left) and 
of the subtracted chiral condensate (right) for various magnetic fluxes 
at $\beta=1.70$ (confinement). Lattice size is 
$16^3 \times 6~$ (FS denotes a finite-size check for $qB=0$ with 
$24^3 \times 6$). Lines show fits with Eq. (3.1).}
\label{fig:fg4}
\end{figure*}
In the left panel of \Fig{fig:fg4} we show the dependence of the bare chiral 
condensate on the quark mass for various values of magnetic flux at $\beta = 1.70$. 
To obtain the results relevant to continuum physics, one has to subtract 
an additive divergence for finite quark mass, as well as do a multiplicative 
renormalization, which is needed also at zero mass.
In the right panel of \Fig{fig:fg4} we show the difference between the bare chiral
condensate for finite fluxes subtracted by the same quantity at zero flux. 
This eliminates the main part of the additive divergence. In the left panel
we have also included points at vanishing quark mass, where there are no additive 
divergencies. The non-vanishing values in this limit are obtained by a chiral 
extrapolation. Because we are not very far from the transition, 
we have supposed a behaviour as for the reduced three-dimensional model 
\cite{Pisarski:1983ms} 
\beq
a^3<\bar{\psi}\psi>  = a_0 + a_1 \sqrt{ma} + a_2 ma. 
\label{eq:3d}
\eeq
Thus, we clearly see chiral symmetry breaking for $\beta=1.70$.
\begin{figure*}[tb]
\vspace*{-2cm}
\includegraphics[width=\textwidth]{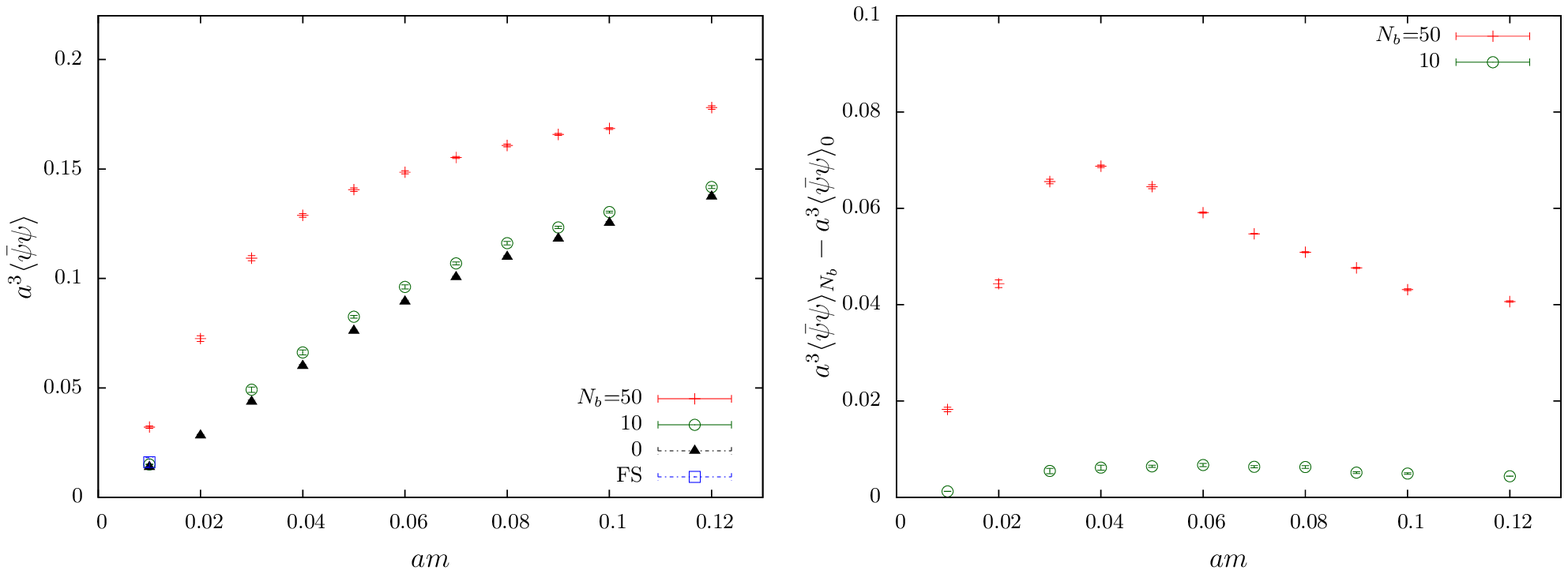}
\vspace*{-15cm}
\caption{Same as for Fig. 4 but in the transition region ($\beta=1.90$).}
\label{fig:fg5}
\end{figure*}
In \Fig{fig:fg5} we show the mass dependence of the bare chiral condensate
(left panel) and the subtracted chiral condensate (right panel) 
in the transition region (at $\beta=1.90$) for three values of the 
magnetic flux. One can see that for finite flux, as well as for zero flux,
the bare and subtracted chiral condensates are consistent with extrapolating 
to zero in the chiral limit. For the highest flux, $N_b=50$ one can clearly
discern two regions of behaviour. For $am \ageq 0.04$ the chiral condensate 
seems to extrapolate to a finite value, but for $am \aleq 0.04$ it actually 
extrapolates to zero. This can be understood, if one assumes that the 
transition for $N_b=50$ at this value of $\beta$ takes place for 
$am\approx 0.04$. 
\begin{figure*}[tb]
\vspace*{-2cm}
\includegraphics[width=\textwidth]{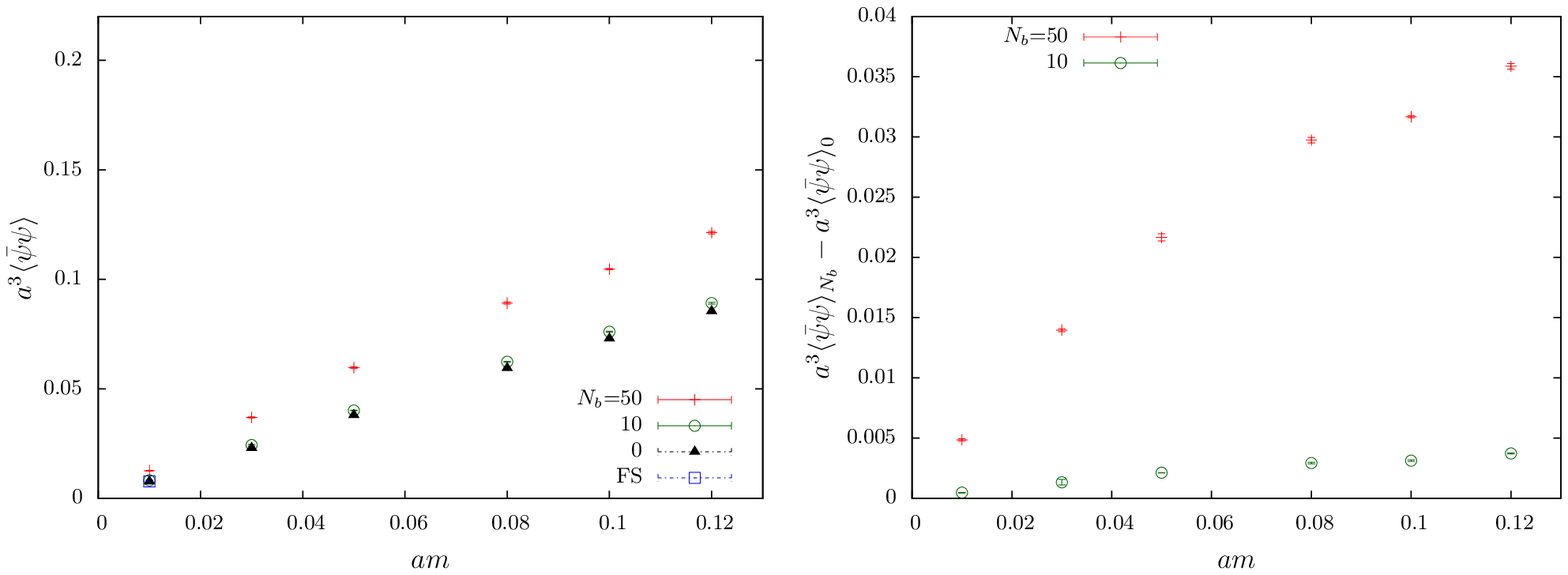}
\vspace*{-15cm}
\caption{Same as for Fig. 4 but in the deconfinement phase ($\beta=2.1$).} 
\label{fig:fg6}
\end{figure*}
In \Fig{fig:fg6} we present the same quantities as above, 
but for $\beta=2.10$. This is well inside the chirally restored phase. 
The chiral condensate extrapolates to zero for all values of the flux.
Thus, chiral symmetry is restored for all values of the flux that we have 
investigated.

In our recent investigation \cite{Ilgenfritz:2013ara} we have used a 
fixed-scale approach, i.e. we kept $\beta$ fixed and thereby the lattice 
spacing $a$ and varied the temperature by changing $\Nt$. In this way
we may easily fix the mass value as well as the magnetic field strength,
while varying the temperature. Moreover, for the time being we may 
neglect renormalization effects.  
More precisely we simulated the theory at $\beta=1.80$ mainly with 
lattice sizes $32^3\times \Nt\,,~\Nt=4,6,8,10$,  and with an even lower mass 
value $ma=0.0025$ (cf. \Tab{tab}) taking each time as a minimum three 
values of the magnetic flux $qB=0.0,~0.67,~1.69~\mathrm{GeV}^2$
corresponding to flux unit numbers $N_b=0, 80, 200$, respectively.

\begin{figure*}[h!t]
\centering
\includegraphics[width=\textwidth]{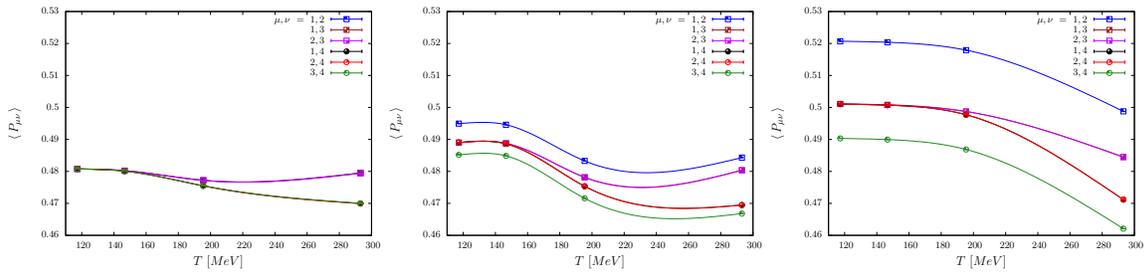}
\caption{Plaquette energies $\langle P_{\mu\nu} \rangle$ vs. 
temperature $T=(a(\beta) \Nt)^{-1}$ 
for $qB=0$ (left), $qB=0.67~\mathrm{GeV}^{2}$ (middle),
and $qB=1.69~\mathrm{GeV}^{2}$ (right) for different plaquette 
orientations ($\beta=1.80, am=0.0025, \Ns=32$). Lines are to
guide the eye.}
\label{fig:fg7}
\end{figure*}

The influence of the magnetic field on the gauge field can be 
represented by studying the different parts $P_{\mu\nu}$ of the gluonic 
action. We introduce variables as in \cite{Bali:2013esa}
\beq
\cE_i^2  =  \langle P_{4i} \rangle, \qquad
\cB_i^2  =  \mid \epsilon_{ijk}\mid \langle P_{jk} \rangle \,, \quad j<k\,.
\eeq
At $B=T=0$ they are all equal by symmetry. At $B=0, T\neq 0$ 
they fall into two groups, because the fourth direction is not 
equivalent to the other ones:
\beq
\cE_1^2 = 
\cE_2^2 \, = \, 
\cE_3^2\, \le \,
\cB_1^2 = 
\cB_2^2 \, = \, 
\cB_3^2\,.    
\eeq
Introducing a magnetic field in the third direction, for $T\neq 0$ 
the only symmetries left are rotations in the $(1,2)$-plane. 
We therefore may define 
\beq
\cE^2_{\parallel}    \equiv   \cE^2_3\,, \qquad
\cE^2_{\perp}        \equiv   \cE^2_1 =\cE^2_2\,, \qquad
\cB^2_{\parallel}    \equiv   \cB^2_3\,, \qquad
\cB^2_{\perp}        \equiv   \cB^2_1 =\cB^2_2\,.
\eeq
In \Fig{fig:fg7} we show the results for the four temperature values, 
and each of them for the three values of the magnetic field. 
We can see the following features from this figure. 
The pattern of the splitting is the same as in our previous
article \cite{Ilgenfritz:2012fw} and more recently found 
in full QCD \cite{Bali:2013esa},
\beq
 \cB^2_{\parallel} \,\, \geq \,\, 
 \cB^2_{\perp}     \,\, \geq \,\,
 \cE^2_{\perp}     \,\, \geq \,\,
 \cE^2_{\parallel} \,\,.
\eeq
Furthermore, if $~\cB^2_{\perp} -\cE^2_{\perp} > 0$ -- which can be 
interpreted as a contribution to the entropy --  
the system can be expected to be at the onset 
of the deconfinement transition or even inside the deconfined phase.
If we compare the middle panel ($qB=0.67~\mathrm{GeV}^2$)
of \Fig{fig:fg7} with the left one ($qB=0$) then at $T=195$ MeV ($\Nt=6$) we find 
this difference to be slightly larger than for the left panel.  
This might be an indication that the transition temperature as a function of 
the temperature went down a bit with increasing magnetic field strength. 
However, comparing the right panel 
($qB=1.69~\mathrm{GeV}^2$) with the left one, then the corresponding difference 
is definitely smaller than for zero magnetic field. This indicates that the 
transition might be shifted to a higher temperature value. 

\begin{figure*}[h!t]
\centering
\includegraphics[width=\textwidth]{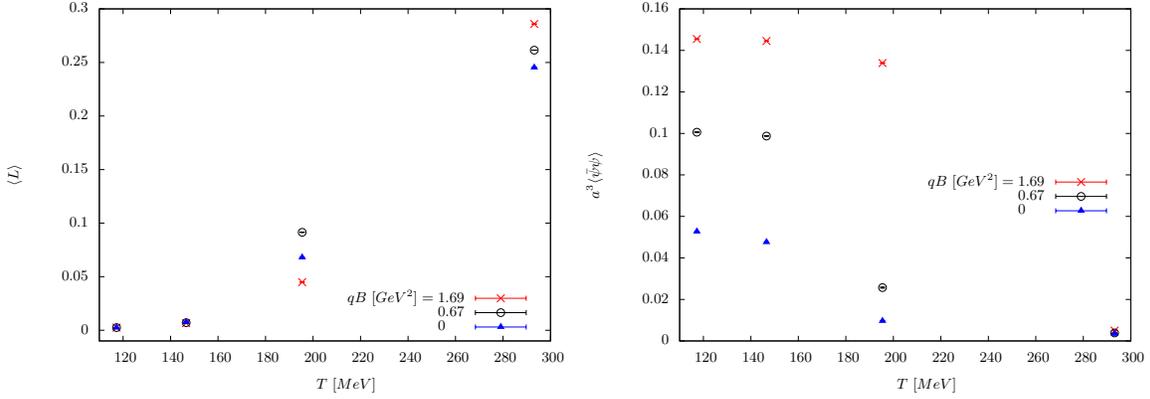}
\caption{Bare Polyakov loop $\langle L \rangle$ (left) and
bare chiral condensate $\langle \bar{\psi}\psi \rangle$ (right)
vs. $T=(a(\beta) \Nt)^{-1}$ for three values of the magnetic field 
strength at $\beta=1.80,~am=0.0025$ and $32^{3} \times \Nt, \; \Nt = 4,6,8,10$.}
\label{fig:fg8}
\end{figure*}
In \Fig{fig:fg8} (left) the expectation value of the unrenormalized Polyakov 
loop $\langle L \rangle$ is shown as a function of the temperature. 
Our sizes $\Nt=4,\ldots,10$ correspond to temperature values $T$, which 
are quite widely spaced. Therefore, we cannot localize the transition e.g. 
for $B=0$ very well. It happens around $T = T_c \simeq 160 - 190$ MeV. 
At $T=195$ MeV ($\Nt=6$) we clearly observe again, that the Polyakov loop does 
not behave monotonously with the magnetic field (see also \cite{D'Elia:2010nq}). 
In \Fig{fig:fg8} (right) the unrenormalized chiral order parameter 
$a^3 \langle \bar{\psi}\psi \rangle$ is shown versus $T$. 
For a fixed non-vanishing quark mass it is increasing monotonously
with the magnetic field at least for the three lower temperature values 
we have investigated. This might mean that $T_c$ always increases with a 
rising magnetic field strength as required for the {\it magnetic catalysis}. 
In particular at $T=195$ MeV we observe a strong increase of the condensate 
between $qB=0.67~\mathrm{GeV}^2$ and our largest value $1.69~\mathrm{GeV}^2$ 
indicating that the system `jumps' from chiral symmetry restoration to the 
chirally broken phase. This indicates that at this temperature value and 
within the given range of magnetic field strength values the critical 
$T_c(B)$ is rising. We find this confirmed, as previously shown in
\Fig{fig:fg5} corresponding to the transition region. At the temperature 
$T=195$ MeV ($\Nt=6$) the chiral extrapolation of the condensate 
$a^3 \langle \bar{\psi}\psi \rangle$ for $qB=0$ and $qB=0.67~\mathrm{GeV}^2$ 
points to zero, i.e. to the chirally restored phase, while for the stronger 
magnetic field strength $1.69~\mathrm{GeV}^2$ the data suggest a non-vanishing 
chiral condensate in the chiral limit (see \cite{Ilgenfritz:2013ara}). 
Thus, we may conclude that at very strong magnetic field values the 
transition temperature grows with $B$. This means {\it magnetic catalysis} 
in agreement with various models \cite{Shovkovy:2012zn}. 

\begin{figure*}[h!t]
\centering
\includegraphics[width=\textwidth]{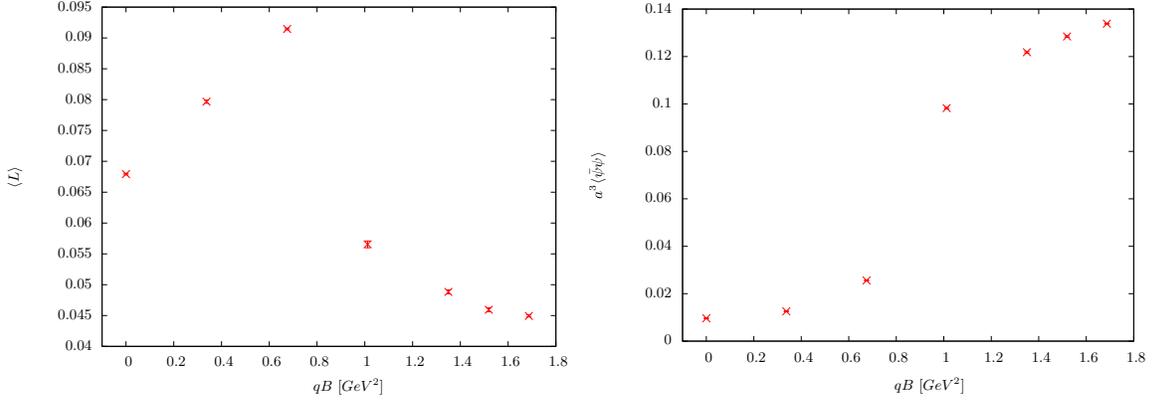}
\caption{Polyakov loop (left panel) and chiral condensate (right panel)
vs. field strength $qB$ at $T=195~\mathrm{MeV}$ 
obtained with $\beta=1.80, am=0.0025$ and $32^3 \times 6$.}
\label{fig:fg9}
\end{figure*}

In order to study the situation in more detail, we have made simulations at 
the same temperature $T=195$ MeV with a few more values of $N_b$. The latter 
correspond to a range of $qB~$ between $0$ and $1.69~\mathrm{GeV}^2$. We 
measure the expectation values of the Polyakov loop and the chiral condensate. 
The results are shown in \Fig{fig:fg9}. 
There is a sharp change, which might be related to a
phase transition in the range $0.7\, \mathrm{GeV}^2 < qB < 1.0~\mathrm{GeV}^2$ 
corresponding to $\sqrt{qB}/T \approx 4.5$. This observation
is supporting a {\it magnetic catalysis} phenomenon.
But for lower magnetic fields we observe a rise of the Polyakov loop with $qB$ 
towards the transition and only then a drop off followed by a monotonous decrease 
at larger field values (compare with our previous non-monotonicity comment 
to \Fig{fig:fg8} (left)).
The rise at low magnetic field values might mean that we are going deeper 
into the deconfinement region, after which the transition brings us back
into the confinement or chirally broken phase. 
The observation of the rise of the Polyakov loop at low magnetic field values 
resembles the pattern discussed in Refs. \cite{Bruckmann:2013oba}, where it
was related to the {\it inverse magnetic catalysis} phenomenon. 

\section{Conclusions}
\label{sec:conclusions} 
\begin{figure}[h!t]
\centering
\includegraphics[width=.45\textwidth]{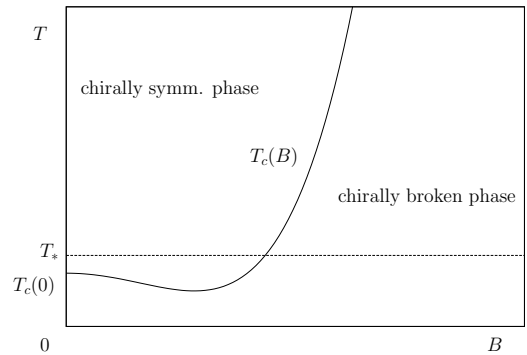}
\caption{Conjectured B-T phase diagram at fixed mass $am=0.0025$. 
The horizontal line $T=T_*=\mathrm{const.}$ indicates the path of 
simulations at $T=195~\mathrm{MeV}$ as in Fig 9.}
\label{fig:fg10}
\end{figure}
Our observations above seem to indicate a decrease of $T_c$ with rising
but small $qB$. At large $qB$ the transition temperature $T_c$ definitely 
rises as expected in the case of a magnetic catalysis. 
In \Fig{fig:fg10} we conjecture a $B-T$ phase 
diagram, which might clarify the situation. In order to prove it, 
further simulations at somewhat smaller temperatures and/or smaller 
quark mass would be helpful. If it proves to be true then one finds
temperature values, where at $qB=0$ the system is in the confinement
(chirally broken) phase. With increasing $qB$ one passes then the 
chirally restored phase, i.e. the deconfinement or chiral transition 
twice, and ends up again in the confinement phase. Along such a
path in the phase diagram the chiral condensate should  
decrease with $qB$ when entering the chirally restored phase.   
This would mean the existence of {\it inverse magnetic catalysis} 
also in two-colour QCD considered throughout this work 
\cite{Ilgenfritz:2012fw,Ilgenfritz:2013ara}.  

\bibliographystyle{utphys}
\providecommand{\href}[2]{#2}\begingroup\raggedright\endgroup

\end{document}